\documentclass[aps,prd,floatfix,groupedaddress,nofootinbib,superscriptaddress,twocolumn,notitlepage]{revtex4-1}

\usepackage{amsmath,amssymb,bm,color,float,graphicx,mathrsfs}
\usepackage[hidelinks]{hyperref}
\hypersetup{colorlinks=true,linkcolor=blue,citecolor=blue,urlcolor=blue}
\usepackage{Macro}
\usepackage{comment}
\usepackage{graphicx}
\usepackage{tikz}
\usepackage{lipsum} % 用于生成示例文本
\usepackage[percent]{overpic}
\usepackage{graphicx}

\begin{document}

% Title %
\title{Impact of reionization history on constraining primordial gravitational waves in future all-sky cosmic microwave background experiments}

\author{Hanchun Jiang}
\affiliation{Department of Physics, Graduate School of Science, The University of Tokyo, Tokyo 113- 0033, Japan}
\author{Toshiya Namikawa}
\affiliation{Department of Applied Mathematics and Theoretical Physics, University of Cambridge, Wilberforce Road, Cambridge CB3 0WA, United Kingdom}
\affiliation{Center for Data-Driven Discovery, Kavli IPMU (WPI), UTIAS, The University of Tokyo, Kashiwa, 277-8583, Japan}
\affiliation{Kavli Institute for Cosmology, Cambridge, Madingley Road, Cambridge CB3 0HA, United Kingdom}

% Date %
\date{\today}

% Abstract %
\begin{abstract}
We explore the impact of the reionization history on examining the shape of the power spectrum of the primordial gravitational waves (PGWs) with the cosmic microwave background (CMB) polarization. The large-scale CMB generated from the reionization epoch is important in probing the PGWs from all-sky experiments, such as LiteBIRD. The reionization model has been constrained by several astrophysical observations. However, its uncertainty could impact constraining models of the PGWs if we use large-scale CMB polarization. Here, by expanding the analysis of Mortonson \& Hu (2007), we estimate how reionization uncertainty impacts constraints on a generic primordial tensor power spectrum. We assume that CMB polarization is measured by a LiteBIRD-like experiment and the tanh model is adopted for a theoretical template when we fit data. We show that constraints are almost unchanged even if the true reionization history is described by an exponential model, where all parameters are within 68\% Confidence Level (CL). We also show an example of the reionization history that the constraints on the PGWs are biased more than 68\% CL. Even in that case, using $E$-mode power spectrum on large scales would exclude such a scenario and make the PGW constraints robust against the reionization uncertainties. 
\end{abstract} 


%////////////////////////////////////////%
% MAIN MATTER 
%////////////////////////////////////////%

% Contents %
\maketitle

%////////////////////////////////////////%
\section{Introduction} \label{sec:intro}
%////////////////////////////////////////%

%Big picture: Inflation->scalar and tensor
Measurements of the cosmic microwave background (CMB) anisotropies have extensively explored the origin of the universe. For example, observed CMB angular power spectra are in excellent agreement with the prediction of the $\Lambda$ Cold-Dark-Matter ($\Lambda$CDM) model. In the standard cosmological scenario, cosmic inflation, a period of rapid quasi-exponential expansion in the early universe \cite{brout1978creation,kazanas1980dynamics,starobinsky1980new,guth1981inflationary,sato1981first,albrecht1982cosmology,linde1982new}, generates primordial density fluctuations through quantum fluctuations in the space-time metric \cite{mukhanov1981quantum,guth1982fluctuations,hawking1982development,linde1982scalar,starobinsky1982dynamics,bardeen1983spontaneous}, which will eventually become the structure of the universe. While the predicted primordial density fluctuations have already been tested by various astrophysical observations, the gravitational waves produced in the very early universe (primordial gravitational waves, PGWs) \cite{starobinskii1979spectrum,rubakov1982graviton,fabbri1983effect,abbott1984constraints} predicted by cosmic inflation remain unconfirmed.

%sub-topic 1: CMB observation / Current observational status of primordial gravitational waves
The most effective method for detecting PGWs is through the curl pattern ($B$-modes) in CMB polarization maps, as linear-order density fluctuations do not generate $B$-modes but are sensitive to PGWs \cite{kamionkowski1997probe,kamionkowski1997statistics,seljak1997measuring,seljak1997signature,zaldarriaga1997all}. The amplitude of the tensor power spectrum parameterized by the tensor-to-scalar ratio, $r$, provides insights into the energy scale of inflation \cite{davis1992cosmic,kamionkowski2016quest,achucarro2022inflation,ade2021improved}. Current CMB observations set upper limits on PGWs, refining inflationary models, with the best constraints from combined BICEP/Keck Array, Planck, and WMAP data, yielding $r<0.032$ at the pivot scale $k_{\rm pivot}=0.05\,$Mpc$^{-1}$ at 95\% Confidence Level (CL) \cite{Tristram:2022}. Currently, the Simons Observatory is observing the $B$-mode polarization precisely to reduce $68\%$ CL uncertainties of $r$ to $0.002$ \cite{Namikawa:2021gyh,SimonsObservatory:2024:r}. In the 2030s, LiteBIRD, a satellite designed for full-sky observations, aims at achieving the uncertainty to $r$ below $0.001$ \cite{litebird2023probing,LiteBIRD:2023:delens}. If current and future CMB experiments can yield an upper limit at $r\sim0.002$, it will constrain any single-field model with a characteristic potential scale larger than the Planck scale \cite{litebird2023probing}. 

%sub-topic 2: Current constraints on reionization history/ reionization -> CMB polarization
Next-generation CMB experiments aim to detect PGWs predicted by many inflationary models through large-scale $B$-mode signals, particularly at the reionization bump. This bump is shaped by Thomson scattering with free electrons during the reionization epoch, where neutral hydrogen was ionized by energetic photons from early luminous sources (see e.g. \cite{Natarajan:2014rra} for a review). The amplitude and the exact shape of the bump are affected by the poorly constrained reionization history, making it crucial to address uncertainties from reionization history in PGW detection.

CMB measurements have offered insights into the reionization epoch \cite{Lewis:2006ym,WMAP7,Planck:2016mks,Heinrich:2016ojb,Hazra:2018eib,Millea:2018bko,Ahn:2020btj,Qin:2020xrg}. For example, Planck 2018 data constrains the optical depth to the CMB to $\tau=0.054\pm 0.007$ ($68\%$ CL), corresponding to a reionization redshift $z_{\rm re}=7.67\pm 0.73$ ($68\%$ CL) \cite{aghanim2020planck}. This constraint is very close to that obtained recently with the Planck Public Release 4 \cite{Tristram:2023:PR4}. However, if we only use small-scale CMB measurements to avoid the uncertainties in the large-scale $E$-modes, recent constraints suggest $\tau=0.080\pm 0.012$ ($68\%$ CL) \cite{giare2024measuring}, indicating challenges in precision. High-redshift quasar and galaxy observations provide additional constraints. Gunn-Peterson troughs \cite{gunn1965density} suggest reionization completed by $z\sim6$ \cite{fan2006survey}, while Lyman-$\alpha$ optical depth fluctuations in quasar spectra \cite{bosman2022hydrogen} and the inferred low mean free path of ionizing photons \cite{gaikwad2023measuring} indicate a later completion around $z\sim5.2$.

%Problems: 1. reionization -> inflationary parameters
Measurements of the reionization history are even less precise, particularly regarding the time and spatial variations at higher redshifts \cite{Dai:2018nce}. The Ly$\alpha$ emission line, sensitive to neutral hydrogen, is a crucial probe \cite{ouchi2020observations}, but Ly$\alpha$ emitters (LAEs) at high redshifts ($z>7$) are exceedingly rare, making it challenging to use these emitters to constrain the reionization history \cite{nakane2024lyalpha}. In contrast, reconstructing reionization history with CMB needs an accurate polarization observation at large-angular scales to precisely measure the reionization bump \cite{Hu:2003gh,Hazra:2018eib,Watts:2020,Sakamoto:2022nth}. These limitations leave a wide range of models viable under current constraints, highlighting significant uncertainties in our understanding of reionization. The uncertainties from the reionization history might affect the accuracy of reconstructing PGWs from polarization signals \cite{Mortonson:2007tb,Lau:2013zea}. Uncertainties in the reionization history could also impact the constraints on cosmology models beyond $\Lambda$CDM~\cite{Paoletti:2020ndu} and on the kinetic Sunayev Zel'dovich effect from reionization \cite{Gorce:2022cvb}. 

% Comparison with previous work
Mortonson \& Hu (2007) \cite{Mortonson:2007tb} explore the impact of the reionization history on the constraints on $r$, assuming Planck or CMBPol experiment, as well as cosmic-variance limited cases. They assume $r=0.3$ or $0.03$ as a fiducial parameter, both of which are almost excluded by the current $B$-mode measurements, so it is important to study again for $r$ consistent with the current observation. Lau et al. (2013) \cite{Lau:2013zea} investigated this problem in partial sky assuming Planck case. Several works have also studied the impacts of patchy reionization on the constraints on $r$ \cite{jain2023framework,mukherjee2019patchy,jain2024disentangling}.

The present paper builds upon previous studies in two significant ways. First, we examine how uncertainties in the reionization history affect constraints on a generic primordial tensor power spectrum (PTPS). Previous studies typically assumed a standard scale-invariant spectrum for PTPS. However, full-sky CMB experiments, such as those targeting the $B$-mode signal, allow us to constrain the deviation of the tensor power spectrum from the standard PTPS parameterized by $r$. By measuring the $B$-mode power spectrum at both the reionization and recombination bumps, we can differentiate between various models of PGWs, such as the SU(2)-axion model \cite{LiteBIRD:2023zmo}, and explore more general forms of the PTPS \cite{hiramatsu2018reconstruction}.
Second, we improve upon the experimental assumptions of previous studies. Earlier works considered CMB experimental configurations with significantly higher polarization noise than what is expected from upcoming missions or a cosmic-variance-limited scenario, making those assumptions unrealistic. In the near future, LiteBIRD will achieve a polarization noise level of approximately $2\mu$K-arcmin by combining all frequency channels. 
LiteBIRD is highly sensitive to $r$, with a projected 68\% CL uncertainty of $\sigma_r \leq 10^{-3}$ \cite{litebird2023probing}. Under such precise measurements, the impact of reionization history on $r$ constraints becomes more critical than in previous experiments. 

% Structure of this paper
This paper is organized as follows. In Sec.~\ref{sec:model}, we describe the PTPS and reionization history we consider in this paper. In Sec.~\ref{sec:method}, we explain our forecast setup and how we derive parameter constraints. In Sec.~\ref{sec:results}, we show how the uncertainties in the reionization model impact the constraint on the PTPS. Sec.~\ref{sec:summary} is devoted to summary and discussion. 

Throughout this paper, we assume the flat $\Lambda$CDM model as our fiducial model. We assume that the Hubble parameter is $H_0=67.81\,$km/s/Mpc and define $h$ as the Hubble parameter divided by $100$ km/s/Mpc. We also assume the physical baryon energy density, $\Omega_{\rm b}h^2=0.02238$, physical CDM energy density, $\Omega_{\rm c}h^2=0.1201$, spectral index of the scalar primordial power spectrum, $n_{\rm s}=0.9660$, and its amplitude, $A_{\rm s}=2.101\times10^{-9}$. These values are consistent with the results from the Planck collaboration \cite{aghanim2020planck}. %\protect\footnotemark. 

%////////////////////////////////////////%
\section{Models} \label{sec:model}
%////////////////////////////////////////%

In this section, we describe the PTPS and reionization models considered in this paper. We modify the public code, {\tt CLASS}, \citep{blas2011cosmic}, to compute the CMB polarization angular power spectra with these models.

\subsection{The primordial tensor power spectrum}

We generalize the standard scale-invariant spectrum by following the steps outlined in Hiramatsu et al. (2018) \cite{hiramatsu2018reconstruction}. Specifically, we characterize the deviation of the PTPS from the standard case as
\begin{align}
    \mathcal{P}_h(k) = 
    \begin{cases} 
        \mathcal{P}_h^{\text{fid}}(k) + \delta P_i & \text{for } k_{i-1} \leq k < k_i \text{ with } 1 \leq i \leq N, \\
        \mathcal{P}_h^{\text{fid}}(k) & \text{for } k < k_0 \text{ and } k_N \leq k, 
    \end{cases}
\end{align}
where $\mathcal{P}_h(k)$ represents the dimensionless amplitude of the PTPS with $\delta P_i$ being constants. Here, $N$ is the number of bins in logarithmic intervals. The values of $k_n$ are given by $k_n=\alpha^n k_0$, where $\alpha$ controls the logarithmic interval. We set $k_0=10^{-4}$ Mpc$^{-1}$, $\alpha=2.04$ and $N=8$, same as the set in Hiramatsu et al. (2018) \cite{hiramatsu2018reconstruction}. In this paper, we assume 
\begin{equation}
    \mathcal{P}_h^{\rm fid}(k) = A_t \exp \left[ n_t \ln\left(\frac{k}{k_{\text{pivot}}}\right) + \alpha_t \left(\ln \left(\frac{k}{k_{\text{pivot}}}\right)\right)^2 \right] 
    \,,
\end{equation}
where $A_t$ is the amplitude of the PTPS, $n_t$ is the tensor spectral index, and $\alpha_t$ is the running of the tensor spectral index. The amplitude is given by $A_t=rA_s$ defined at the pivot scale, $k_{\rm pivot}=0.05\,\text{Mpc}^{-1}$, where $A_s$ is the amplitude of the curvature perturbations and $r$ is the tensor-to-scalar ratio. We assume that $n_t$ and $\alpha_t$ follow the slow-roll inflationary consistency relation as \cite{lidsey1997reconstructing}:
\begin{align}
    \begin{aligned}
        %-r/8*(2-r/8-n_s)
        n_t&=-\frac{r}{8}\left(2-\frac{r}{8}-n_s\right) \,, \\
        %r/8(r/8+n_s-1)
        \alpha_t&=\frac{r}{8}\left(\frac{r}{8}+n_s-1\right) \,,
    \end{aligned}
\end{align}
where $n_s$ is the scalar spectral index. The PTPS at low- and high-$k$, could be modified in the SU(2)-axion model, while the spectrum is enhanced at intermediate scales by the massive gravity inflation model predicts \cite{hiramatsu2018reconstruction}.

\subsection{Reionization models}

%=========fig start================%
\begin{figure}[t]
    \centering
    \includegraphics[width=\linewidth]{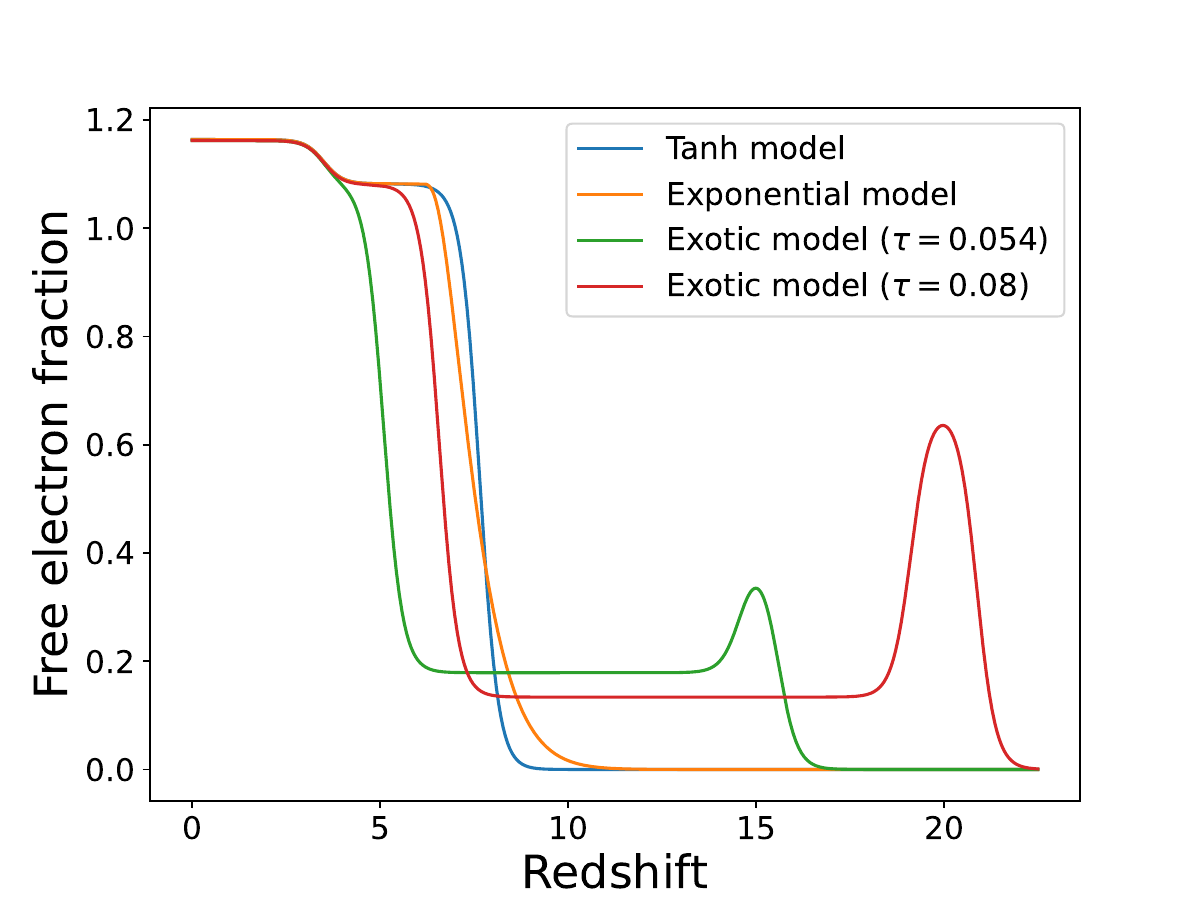}
    \caption{Free electron fraction, $x_{\rm e}$, as a function of redshift for the tanh model (blue line), exponential model (orange line), and an example for exotic models (green line). Here, we set the optical depth as $\tau=0.054$. We also show an example for exotic models with $\tau=0.08$ for comparison (red line).
    }
    \label{fig:xe}
\end{figure}
%===========fig end===============%

We first introduce the hyperbolic tangent (tanh) model that has been widely used in constraining cosmological parameters with CMB angular power spectra. In this model, the reionization history is divided into two phases: the phase before reionization begins ($z \geq z_{\rm start}$), where the free-electron fraction $x_e$ remains constant at $x_{e, \text{before}}$, and reionization epoch ($z<z_{\rm start}$), during which $x_e(z)$ evolves as  
%To describe the reionization history, we consider three epochs; before the reionization starts ($z\geq z_{\rm start}$), the reionization epoch ($z_{\rm c}\leq z<z_{\rm start}$), and after the Hydrogen reionization completed ($z<z_{\rm c}$). For the reionization epoch, the so-called tanh model is often used to describe the reionization history when computing the CMB angular power spectra in data analysis and forecast study. In this model, the evolution of $x_{\rm e}(z)$ is given by

%=========equation start============%
\al{
    x_{\rm e}(z) = (1-x_{e,\rm before})x_{\rm reio}(z) + x_{e,\rm before} + x_{\rm e, He}(z) 
    \,, \label{eq:tanh}
}
%===========equation end============%
where we define
\al{
    x_{\rm reio}(z) = \frac{1}{2}\left[1+\tanh\left(\frac{z_{\rm reio}-z}{\Delta z}\right)\right]
    \,. 
}
Here, $z_{\rm reio}$ is the redshift when $x_{\rm reio}(z)=0.5$ and $\Delta z$ is the duration of the reionization epoch. We fix $\Delta z=0.5$ and vary $z_{\rm reio}$ in our forecast. 
The free-electron fraction before the reionization, $x_{\rm e, before}$, and the contributions from Helium, $x_{\rm e,He}$, are derived from the {\tt HyRec} code implemented in {\tt CLASS}. 
%For simplicity, we assume that the redshift when the reionization is completed, $z_{\rm c}$, is fixed to $6.1$, while the redshift when the reionization starts, 
The starting redshift of reionization, $z_{\rm start}$, is given by $z_{\rm start} = 8.0 \times \Delta z + z_{\rm reio}$.

From the reionization history, $x_e(z)$, we often compute the optical depth to the CMB defined as 
%==========equaiton start============%
\begin{equation}
    \tau = \int_{0}^{\infty} \sigma_{\rm T} \,a(z) \, n_{\rm p} \, x_{\rm e}(z) \, \frac{1}{H(z)(1+z)} \, dz
    \,,
\end{equation}
%===========equation end===========%
where $a$ is the scale factor, $n_{\rm p}$ is proton number density, $\sigma_{\rm T}$ is the Thomson scattering cross-section, and $H(z)=\dot{a}/a$ with the dot denoting the time derivative. We compute the proton number density as $n_{\rm p}=\Omega_{\rm b}h^2\rho_{\rm cr}/m_{\rm p}h^2$ where $\rho_{\rm cr}$ is the critical density, $m_{\rm p}$ is proton mass. 

In our forecast, we consider two different models for the ``true'' reionization history as described in detail in the following subsection and use the tanh model for the theoretical template to fit. Figure \ref{fig:xe} summarizes the reionization history we used in our forecast study. 

\subsubsection{Exponential model}

The first model we consider is the exponential model described in the {\tt CAMB} package \citep{lewis2000efficient} where $x_{\rm reio}(z)$ in Eq.~\eqref{eq:tanh} is replaced to 
\footnote{The definition can be found from the source code of the {\tt CAMB} package: \url{https://github.com/cmbant/CAMB/blob/master/fortran/reionization.f90 } (line 380-403).}
\al{
    x_{\rm reio}(z) = \exp\left[-\lambda \frac{(z-z_{\rm c})^{3/2}}{1+[\Delta z/(z-z_{\rm c})^2]}\right] 
    \,.
}
Here, the evolution rate in the exponential, $\lambda$, is defined as
\begin{equation}
    \lambda = \frac{-\ln0.5}{(z_*-z_{\rm c})^{2/3}}
    \,. \label{eq:lambda}
\end{equation}
For simplicity, we assume that the redshift when the reionization is completed, $z_{\rm c}$, is fixed to $6.1$. We set $z_*=7.249$, 
%and $z_{\rm start}=8.0\times\Delta z+z_*$, 
which corresponds to $\tau = 0.054$.

\subsubsection{Exotic reionization model}

%=========fig start================%
\begin{figure*}[t]
    \centering
    \includegraphics[width=\linewidth]{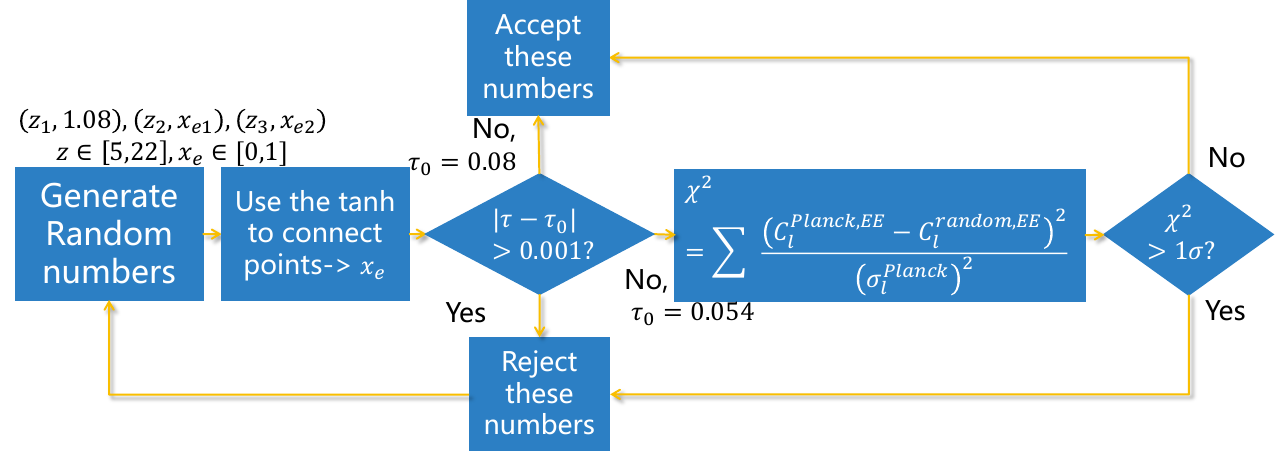}
    \caption{The process to generate random $x_{\rm e}$ model. We generate three random redshift points within $5\sim22$, among them $z_1$ is the end of the reionization and $x_{\rm e}(z_1)$ is 1.08 which corresponds to the Helium reionization. We also generate two random numbers $x_{\rm e}(z_2)$ and $x_{\rm e}(z_3)$ within $0\sim1$. After that, we use the tanh function to connect them. We consider two cases: $\tau=0.054$ and $\tau=0.08$. To generate a reionization history with $\tau=0.08$, we will accept these numbers if $|\tau-\tau_0|<0.001$. For $\tau=0.054$, we will also verify this model consistent with Planck's result \cite{tristram2023cosmological}.}
	\label{fig:random}
\end{figure*}
%===========fig start===============%

TThe other model we consider for the true reionization history is an exotic model. It demonstrates how the PTPS constraint depends on the assumed reionization history. We generate the parametrized reionization history as a function of $z$ using random points. The process is shown in Fig.~\ref{fig:random}. Note that we generate many exotic models, but this paper only shows the forecast results for a model that introduces the most significant bias among them. 

In the process of generating random $x_e$ models, we generate three random redshift points within the range $5\leq z\leq 22$. Among these, $z_1$ marks the end of reionization, where $x_e(z_1)=1.08$, corresponding to helium reionization. We also generate two random numbers, $x_e(z_2)$ and $x_e(z_3)$, between 0 and 1. A tanh function is used to connect these points. We then obtain $\tau$ from this reionization history and reject the model if $|\tau-\tau_0|>0.001$ where $\tau_0$ is either $0.054$ or $0.08$ motivated by the recent observations of Ref.~\cite{tristram2023cosmological} and \cite{giare2024measuring}, respectively. For $\tau=0.054$, we reject models which deviate significantly from the large-scale Planck $E$-mode power spectrum. Specifically, we compute $\chi^2\equiv \sum_{l=2}^{30}(C_l^{EE,{\rm Planck}}-C_l^{EE,{\rm random}})^2/\sigma^2_l$ where $C_l^{EE,{\rm Planck}}$ is the Planck PR4 $E$-mode power spectrum, $C_l^{EE,{\rm random}}$ is the $E$-mode power spectrum with a generated $x_{\rm e}(z)$, and $\sigma_l$ is the measurement error of Planck PR4 data taken from Ref.~\cite{tristram2023cosmological}. For $\tau_0=0.08$, we do not consider further rejections since $\tau=0.08$ is obtained without relying on the large-scale Planck $E$-mode power spectrum \cite{giare2024measuring}. 
After randomly generating $x_e$ models, we choose a case where the reionization history significantly deviates from the tanh case. We generate the models for $\tau_0=0.054$ and $\tau_0=0.08$ respectively. The selected reionization history exhibits behavior similar to the double reionization model \cite{cen2003universe}.

%////////////////////////////////////////%
\section{Method} \label{sec:method}
%////////////////////////////////////////%

Here, we describe our method for the forecast study. We jointly constrain the reionization history and PTPS with the $E$- and $B$-mode power spectra with a Markov Chain Monte Carlo (MCMC) analysis \cite{kosowsky2002efficient}. Our assumptions on the CMB experimental configurations and likelihood are described as follows. 

We adopt a LiteBIRD-like CMB experiment, assuming that the observed polarization data has white noise and is convolved with a Gaussian circular beam for simplicity. The noise power spectrum for the beam-deconvolved CMB polarization map is given by (e.g. Ref.~\cite{namikawa2010probing})
%============equation start========%
\begin{equation}
    N_{l} = \left(\frac{\sigma}{T_{\rm CMB}}\frac{\pi}{10800}\right)^2\exp\left[\frac{l(l+1)}{8\ln2}\left(\theta\frac{\pi}{10800}\right)^2\right]
    \,,
\end{equation}
%===========equation end============% 
where $T_{\rm CMB}=2.75$\,K is the CMB black-body temperature, $\theta$ is the Full Width at Half Maximum (FWHM) of the circular beam in arcmin, and $\sigma$ is the noise level of the polarization map in $\mu$K-arcmin. For LiteBIRD, we assume $\theta=30$ arcmin and $\sigma=2\mu$K-arcmin \cite{hiramatsu2018reconstruction,LiteBIRD:2023zmo}.

The likelihood for the $E$- and $B$-mode power spectra in an idealistic full-sky observation are given as a Wishart distribution \cite{Hamimeche:2008ai}. Specifically, we employ the following log-likelihood function \cite{katayama2011simple,litebird2023probing}:
\begin{widetext}
%==========equaiton start============%
\al{
    -2\ln\mC{L}(\bm{p}) = \sum_{X=E,B}\sum_{l=l_{\rm min}}^{l_{\rm max}} 
    f_{\rm sky}(2l+1)\left[\frac{C_l^{XX,\rm fid}+N_l}{C_l^{XX}(\bm{p})+N_l}+\ln(C_l^{XX}(\bm{p})+N_l)-\frac{2l-1}{2l+1}\ln(C_l^{XX,\rm fid}+N_l)\right]
    \,. \label{Eq:lnL}
}
%===========equation end===========%
\end{widetext}
Here, $C^{XX,\rm fid}_l$ represents mock data and is computed either with the exponential or exotic reionization history. 
The theoretical power spectra, $C_l^{XX}(\bm{p})$, are defined using the tanh function to describe the reionization history, and the vector $\bm{p}$ refers to the set of free parameters.
The polarization noise power spectrum is denoted by $N_l$.  
The minimum and maximum multipoles of the power spectra are $l_{\rm min}$ and $l_{\rm max}$, respectively, where $l_{\rm min}$ is set to $2$. For the maximum multipole, we choose $l_{\rm max}=1300$ for the $E$-mode power spectrum and $1100$ for the $B$-mode power spectrum. Our results are not sensitive to the maximum multipole of the $B$-mode power spectrum as long as we set $l_{\rm max}\gg 100$ where the PGW contributions are not important. 
Finally, $f_{\rm sky}$ is the sky coverage, for which we set $f_{\rm sky}=0.7$, and we follow Ref.~\cite{LiteBIRD:2023zmo} where the entire log-likelihood is scaled by $f_{\rm sky}$. 

We consider $r$, $\delta P_i$'s, and the optical depth to the CMB, $\tau$, as the free parameters. \footnote{In our calculation, we vary $z_{\rm reio}$ in the tanh model. Since we fix $\Omega_{\bm b}h^2$, we obtain $\tau$ from $z_{\rm reio}$.} 
Since the small-scale $E$-mode power spectrum, given by $A_s e^{-2\tau}$, is well constrained by the temperature power spectrum, we fix $A_s e^{-2\tau}$ by adjusting $A_s$ to keep it unchanged \cite{Mortonson:2007tb}. For the fiducial parameters, we test two cases: $r = 0.01$ and $r = 0.001$, while setting $\delta P_i = 0$. For the optical depth, we consider $\tau=0.054$ that matches the latest result from the Planck collaboration \cite{aghanim2020planck}, and $0.080$ obtained without relying on the large-scale $E$-mode measurement \cite{giare2024measuring}. 
\footnotetext{\url{https://wiki.cosmos.esa.int/planck-legacy-archive/images/b/be/Baseline_params_table_2018_68pc.pdf}}
We modify {\tt emcee} \cite{foreman2013emcee} to perform the MCMC analysis. 
We also modify {\tt CLASS} \cite{lewis2000efficient} to compute the power spectra with the reionization histories and PTPS described above.

%////////////////////////////////////////%
\section{Results} \label{sec:results}
%////////////////////////////////////////%

%=========fig end================%
\begin{figure*}[t]
    \centering
    \includegraphics[width=1\linewidth]{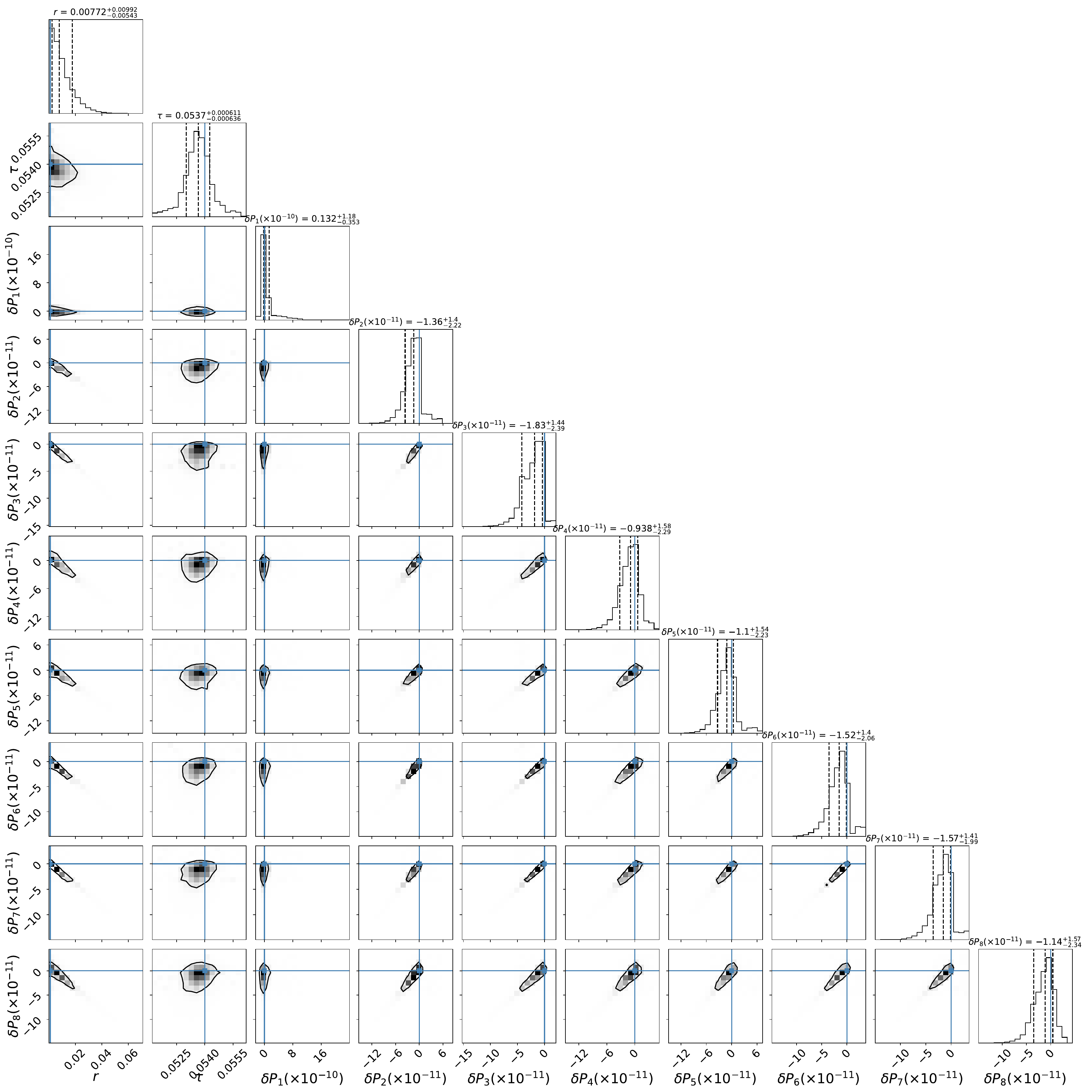}
    \caption{The posterior distribution of parameters for the exponential model case, with $r=0.001$, $\tau=0.054$, and $\delta P_i=0$. The blue lines show the fiducial values. The dashed lines in the 1D posterior histogram represent the $16\%$, $50\%$, and $84\%$ percentiles of the sample, from left to right. The contours in the 2D posterior represent the $68\%$ confidence region ($1\sigma$) of the parameter estimates. 
    }
\label{fig:exp_triangle}
\end{figure*}
%===========fig end===============%

%=========fig start================%
\begin{figure*}[t]
    \centering
    \includegraphics[width=\linewidth]{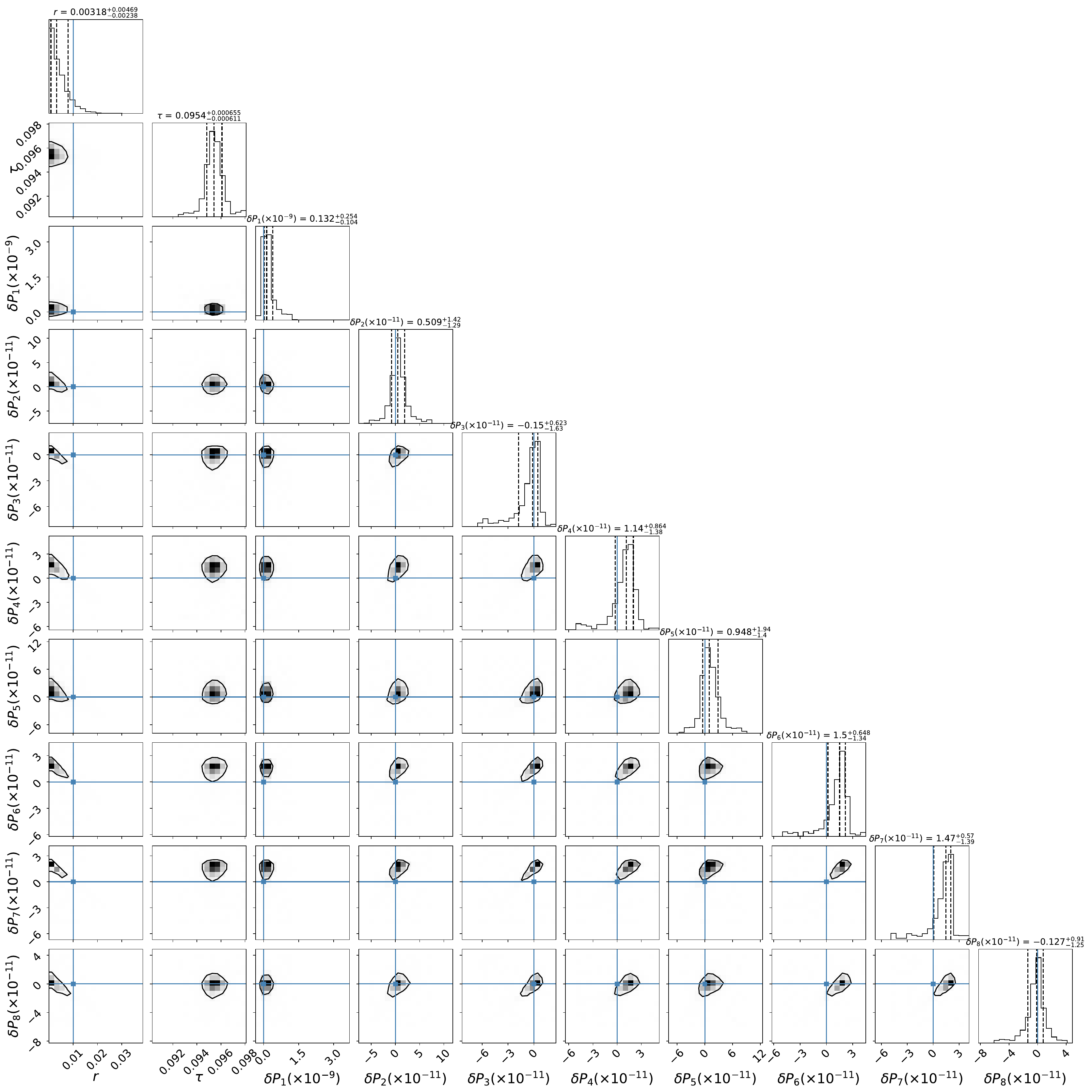}
    \caption{Same as Fig.~\ref{fig:exp_triangle} but for the exotic model with $r=0.01$, $\tau=0.08$, and $\delta P_i=0$.}
    \label{fig:random_triangle}
\end{figure*}
%===========fig end===============%

%=========fig start================%
\begin{figure}[t] 
    \centering
    \begin{overpic}[width=\linewidth]{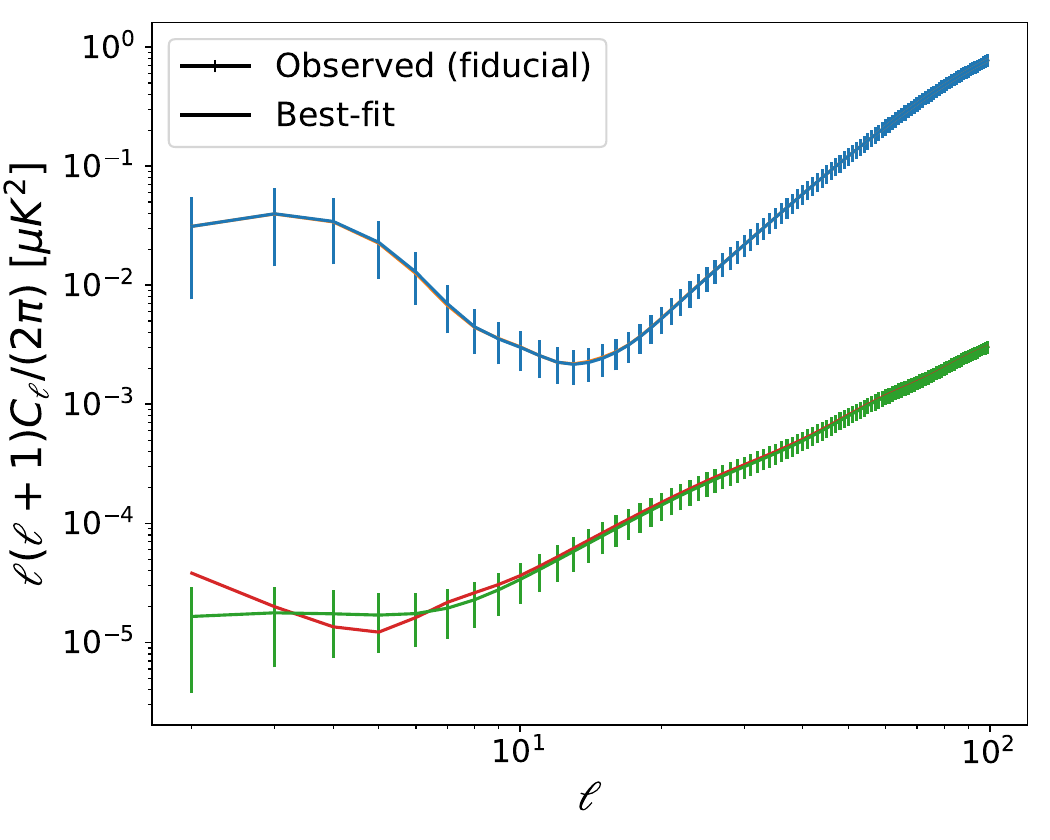} 
        \put(40,55){$C_l^{EE}$}
        \put(40,26){$C_l^{BB}$}
    \end{overpic}
    \caption{$E$- and $B$-mode power spectra with the best-fit values and with the fiducial values with the $1\sigma$ observational errors for the exponential model.}
    \label{fig:exp_ps}
\end{figure}
%=========fig end================%

%=========fig start================%
\begin{figure}[t] 
    \centering
    \begin{overpic}[width=\linewidth]{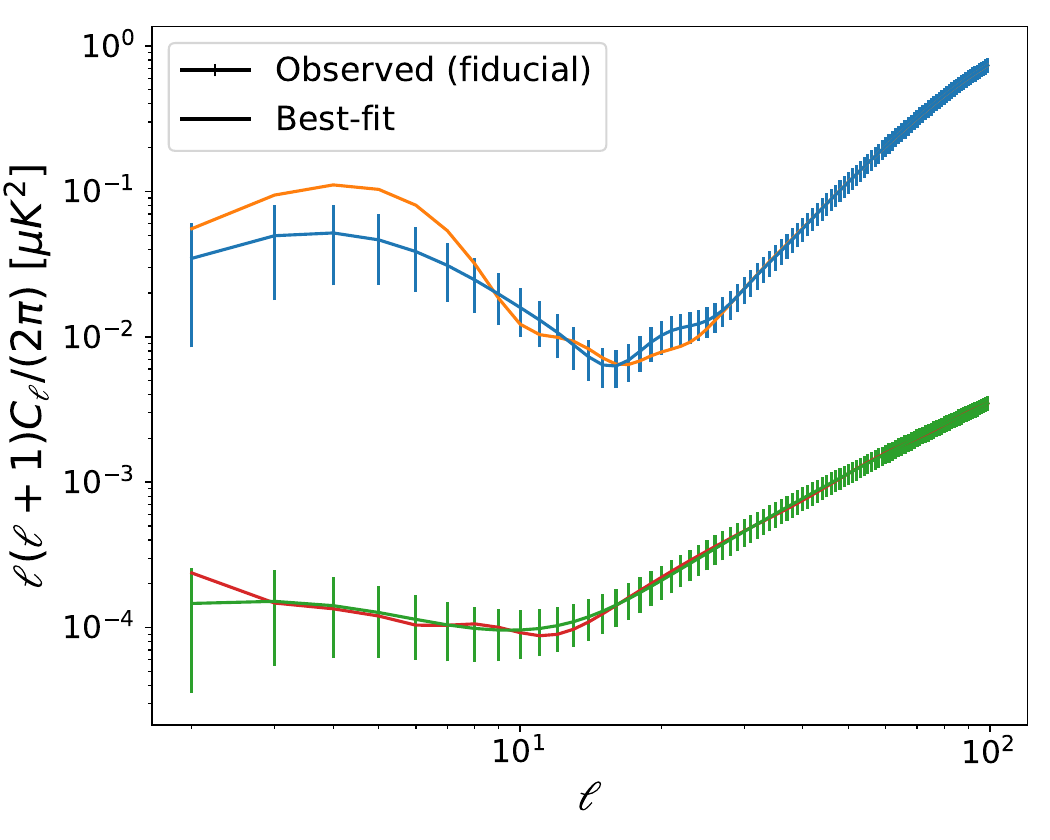}
        \put(40,60){$C_l^{EE}$}
        \put(40,24){$C_l^{BB}$}
    \end{overpic}
    \caption{Same as Fig.~\ref{fig:exp_ps} but for the exotic model.}
    \label{fig:random_ps}
\end{figure}
%=========fig end================%

In this section, we show the results of our forecast. We use the tanh model to fit the mock power spectra in which the exponential or exotic models are used for the reionization history. 

Figure \ref{fig:exp_triangle} shows the results of the MCMC analysis where the fiducial model is the exponential model with $r=0.001$, $\tau=0.054$, and $\delta P_i=0$. The fiducial values are within the $68\%$ confidence contours, showing that the incorrect tanh reionization model did not result in a significant bias to the constraints on the PTPS. Note that we also change the optical depth to $\tau=0.08$ and the tensor-to-scalar ratio to $r=0.01$, but find that the parameters are also within 68\% confidence contours.
The constraint at $68\%$ CL is approximately consistent with that obtained in Ref.~\cite{hiramatsu2018reconstruction}, although there are several differences in the forecast setup. 
We find that $\delta P_i$'s except for $\delta P_1$ show a positive degeneracy. Conversely, $r$ shows negative degeneracy with $\delta P_i$ other than $\delta P_1$. This behavior is attributed to the fact that an increase (decrease) of $r$ is compensated by decreasing (increasing) $\delta P_i$ with $i\geq 2$. In contrast, if $\delta P_1$ increases, the $B$-mode power spectrum at $l=2$ is enhanced while $l=3$ is enhanced slightly, leading to a scale-dependent change in the power spectrum. The optical depth is almost not degenerated with other parameters as the reionization history would be tightly constrained by the reionization bump in the $E$-mode power spectrum. 

Figure \ref{fig:random_triangle} shows the results of the parameter constraints for the exotic model whose reionization history, $x_{\rm e}(z)$, is shown as the red line in Fig.~\ref{fig:xe}. The fiducial values of the parameters are $r=0.01$, $\tau=0.08$, and $\delta P_i=0$. The results show that the best-fit values are biased by more than $1\sigma$, especially for the tensor-to-scalar ratio, optical depth, $\delta P_6$ and $\delta P_7$. 
The bias to the small-scale tensor amplitudes, $\delta P_6$ and $\delta P_7$, is the result of the change in the $B$-mode power spectrum at high-$l$ by the exotic model; the high ionization fraction at high redshift amplifies the polarization power spectra at multipoles larger than the reionization bump \cite{Hu:2003gh}. At $z\sim 20$, this amplification affects both the $E$- and $B$-mode power spectra at higher multipoles, i.e., $l=10-30$, than the reionization bump. 
To fit the amplified reionization bump in the $B$-mode power spectrum, the best-fit $r$ becomes very small, leading to a significant bias in the value of $r$.

Figures \ref{fig:exp_ps} and \ref{fig:random_ps} show the corresponding $E$- and $B$-mode power spectra with the best-fit values for the exponential and exotic models, respectively, and compare them with the fiducial power spectra with the observational errors at $68\%$ CL at each multipole for the LiteBIRD-like experiment. The observational errors at each multipole on the $E$- and $B$-mode power spectra are given by (e.g., Ref.~\cite{namikawa2010probing}):
%=======equation start=========%
\al{
    \Delta C_l^{XX} = \frac{C^{XX,{\rm fid}}_l+N_l}{\sqrt{(l+1/2)f_{\rm sky}}}
    \,.
}
%=======equation end===========%
For the exponential model, both the best-fit $E$- and $B$-mode power spectra show excellent agreement with the fiducial spectrum. However, for the exotic model, while the $B$-mode power spectrum can be fitted well, the best-fit $E$-mode power spectrum is far from the fiducial power spectrum, especially for the reionization bump. This discrepancy in the $E$-mode power spectrum leads to a significant bias in $\tau$ shown in Fig.~\ref{fig:random_triangle}.

%////////////////////////////////////////%
\section{Summary and Discussion} \label{sec:summary}
%////////////////////////////////////////%

We explored how an incorrect model of the reionization history impacts the constraints on the PTPS for the LiteBIRD-like experiment. In the exponential model case, we found that the mock $E$- and $B$-mode power spectra agree well with that using the best-fit parameters within the observational errors. 
In the exotic scenario, we showed that $r$ and $\delta P_i$ at small scales are biased. However, the $E$-mode power spectrum is not in good agreement with that computed with the best-fit values and such a scenario would be easily excluded by measuring the large-scale $E$-mode power spectrum. 
%Our results showed that the constraint on the tensor primordial power spectrum is not sensitive to the uncertainties in the model of the reionization history if we use both $E$- and $B$-mode power spectra. 
Our results show that incorrect reionization models can introduce significant bias in the constraints on the tensor primordial power spectrum.
To avoid potential bias from incorrect modeling of the reionization history, in future CMB experiments, it is crucial to accurately measure the $E$-mode power spectrum to robustly constrain the PTPS. 

We note that our analysis used several simplifications that do not include any practical issues in measuring the large-scale polarization power spectra. 
% foregrounds
For example, the large-scale polarization is dominated by the Galactic foregrounds. Multiple works have developed foreground-cleaning methods to suppress a bias from the Galactic foregrounds in the $B$-mode power spectrum (e.g. Refs.~\cite{Stompor:2008:FGbuster,katayama2011simple,Ichiki:2018:delta-map,Remazeilles:2020:cMILC,Carones:2022:FG}). Compared to the $B$-mode power spectrum, however, the $E$-mode power spectrum has a large signal, and the bias from foreground residuals would be much smaller than that in the $B$-mode power spectrum. 
% noise
The large-scale multipoles could also be contaminated by $1/f$ noise, although its impact on polarization is much less significant than in temperature. 
% likelihood
The likelihood approximation given in Eq.~\eqref{Eq:lnL} which has been used for multiple forecast studies is not valid if we work on real cut-sky data that removed the Galactic plane and point-source contributions. Instead of introducing the scaling factor, $f_{\rm sky}$, we should use the likelihood approximation proposed by Ref.~\cite{Hamimeche:2008ai} or the exact but computationally-intensive pixel-based likelihood \cite{Planck:2015,Gerbino:2019:likelihood}. 
% delensing
We did not consider the $B$-mode delensing in our forecast. The delensing improves the constraint on $r$ by approximately $50\%$ in the Simons Observatory as the constraint on $r$ is determined by the recombination bump in the $B$-mode power spectrum \cite{SimonsObservatory:2024:r}. For LiteBIRD, the delensing improves the constraint by approximately $10\%$ which is mostly attributed to a reduction of the statistical errors at the recombination bump \cite{Namikawa:2021gyh}. We expect that the delensing does not significantly impact our results as the reionization history modifies mostly the reionization bump. 
A study on these practical issues is beyond our scope and is left for our future work.

%////////////////////////////////////////%
% BACK MATTER 
%////////////////////////////////////////%

% Appendix %
\appendix

\begin{acknowledgments}
We thank Gill Holder and Naoki Yoshida for the useful discussion and helpful comments. TN also thanks Antony Lewis for initiating this project. 
HJ is supported by the International Graduate Program for Excellence in Earth-Space Science (IGPEES). 
TN is supported in part by JSPS KAKENHI Grant No. JP20H05859 and No. JP22K03682. 
The Kavli IPMU is supported by the World Premier International Research Center Initiative (WPI Initiative), MEXT, Japan. 
\end{acknowledgments}

% References %
%\bibliographystyle{apsrev}
\bibliographystyle{mybst}
\bibliography{ref}

\providecommand{\href}[2]{#2}\begingroup\raggedright\begin{thebibliography}{10}

\bibitem{brout1978creation}
R.~Brout, F.~Englert, and E.~Gunzig {\em Annals of Physics} {\bf 115} (1978), no.~1 78--106.

\bibitem{kazanas1980dynamics}
D.~Kazanas {\em Astrophysical Journal, Part 2-Letters to the Editor} {\bf 241} (1980) L59--L63.

\bibitem{starobinsky1980new}
A.~A. Starobinsky {\em \plb} {\bf 91} (1980), no.~1 99--102.

\bibitem{guth1981inflationary}
A.~H. Guth {\em \prd} {\bf 23} (1981), no.~2 347.

\bibitem{sato1981first}
K.~Sato {\em \mnras} {\bf 195} (1981), no.~3 467--479.

\bibitem{albrecht1982cosmology}
A.~Albrecht and P.~J. Steinhardt {\em \prl} {\bf 48} (1982), no.~17 1220.

\bibitem{linde1982new}
A.~D. Linde {\em \plb} {\bf 108} (1982), no.~6 389--393.

\bibitem{mukhanov1981quantum}
V.~F. Mukhanov and G.~Chibisov {\em ZhETF Pisma Redaktsiiu} {\bf 33} (1981) 549--553.

\bibitem{guth1982fluctuations}
A.~H. Guth and S.-Y. Pi {\em \prl} {\bf 49} (1982), no.~15 1110.

\bibitem{hawking1982development}
S.~W. Hawking {\em \plb} {\bf 115} (1982), no.~4 295--297.

\bibitem{linde1982scalar}
A.~D. Linde {\em \plb} {\bf 116} (1982), no.~5 335--339.

\bibitem{starobinsky1982dynamics}
A.~A. Starobinsky {\em \plb} {\bf 117} (1982), no.~3-4 175--178.

\bibitem{bardeen1983spontaneous}
J.~M. Bardeen, P.~J. Steinhardt, and M.~S. Turner {\em \prd} {\bf 28} (1983), no.~4 679.

\bibitem{starobinskii1979spectrum}
A.~Starobinskii {\em JETP Letters} {\bf 30} (1979), no.~11 682--685.

\bibitem{rubakov1982graviton}
V.~Rubakov, M.~V. Sazhin, and A.~Veryaskin {\em \plb} {\bf 115} (1982), no.~3 189--192.

\bibitem{fabbri1983effect}
R.~Fabbri and M.~Pollock {\em \plb} {\bf 125} (1983), no.~6 445--448.

\bibitem{abbott1984constraints}
L.~F. Abbott and M.~B. Wise {\em Nuclear physics B} {\bf 244} (1984), no.~2 541--548.

\bibitem{kamionkowski1997probe}
M.~Kamionkowski, A.~Kosowsky, and A.~Stebbins {\em \prl} {\bf 78} (1997), no.~11 2058.

\bibitem{kamionkowski1997statistics}
M.~Kamionkowski, A.~Kosowsky, and A.~Stebbins {\em \prd} {\bf 55} (1997), no.~12 7368.

\bibitem{seljak1997measuring}
U.~Seljak {\em \apj} {\bf 482} (1997), no.~1 6.

\bibitem{seljak1997signature}
U.~Seljak and M.~Zaldarriaga {\em \prl} {\bf 78} (1997), no.~11 2054.

\bibitem{zaldarriaga1997all}
M.~Zaldarriaga and U.~Seljak {\em \prd} {\bf 55} (1997), no.~4 1830.

\bibitem{davis1992cosmic}
R.~L. Davis, H.~M. Hodges, G.~F. Smoot, P.~J. Steinhardt, and M.~S. Turner {\em \prl} {\bf 69} (1992), no.~13 1856.

\bibitem{kamionkowski2016quest}
M.~Kamionkowski and E.~D. Kovetz {\em Annual Review of Astronomy and Astrophysics} {\bf 54} (2016) 227--269.

\bibitem{achucarro2022inflation}
A.~Ach\'ucarro {\em et~al.} \href{http://arxiv.org/abs/2203.08128}{{\tt arXiv:2203.08128}}.

\bibitem{ade2021improved}
P.~A. Ade, Z.~Ahmed, M.~Amiri, D.~Barkats, R.~B. Thakur, C.~Bischoff, D.~Beck, J.~Bock, H.~Boenish, E.~Bullock, {\em et~al.} {\em \prl} {\bf 127} (2021), no.~15 151301.

\bibitem{Tristram:2022}
M.~Tristram {\em et~al.} {\em \prd} {\bf 105} (2022), no.~8 083524, \href{http://arxiv.org/abs/2112.07961}{{\tt arXiv:2112.07961}}.

\bibitem{Namikawa:2021gyh}
T.~Namikawa {\em et~al.} {\em Phys. Rev. D} {\bf 105} (2022), no.~2 023511, \href{http://arxiv.org/abs/2110.09730}{{\tt arXiv:2110.09730}}.

\bibitem{SimonsObservatory:2024:r}
{\bf Simons Observatory}  {\bf Collaboration} , E.~Hertig {\em et~al.} {\em Phys. Rev. D} {\bf 110} (2024), no.~4 043532, \href{http://arxiv.org/abs/2405.01621}{{\tt arXiv:2405.01621}}.

\bibitem{litebird2023probing}
{LiteBIRD Collaboration}, E.~Allys, K.~Arnold, J.~Aumont, R.~Aurlien, S.~Azzoni, C.~Baccigalupi, A.~Banday, R.~Banerji, R.~Barreiro, {\em et~al.} {\em \ptep} {\bf 2023} (2023), no.~4 042F01, \href{http://arxiv.org/abs/2202.02773}{{\tt arXiv:2202.02773}}.

\bibitem{LiteBIRD:2023:delens}
{\bf LiteBIRD}  {\bf Collaboration} , T.~Namikawa {\em et~al.} {\em JCAP} {\bf 06} (2024) 010, \href{http://arxiv.org/abs/2312.05194}{{\tt arXiv:2312.05194}}.

\bibitem{Natarajan:2014rra}
A.~Natarajan and N.~Yoshida {\em \ptep} {\bf 2014} (2014), no.~6 06B112, \href{http://arxiv.org/abs/1404.7146}{{\tt arXiv:1404.7146}}.

\bibitem{Lewis:2006ym}
A.~Lewis, J.~Weller, and R.~Battye {\em \mnras} {\bf 373} (2006) 561--570, \href{http://arxiv.org/abs/astro-ph/0606552}{{\tt astro-ph/0606552}}.

\bibitem{WMAP7}
E.~{Komatsu}, K.~M. {Smith}, J.~{Dunkley}, C.~L. {Bennett}, B.~{Gold}, G.~{Hinshaw}, N.~{Jarosik}, D.~{Larson}, M.~R. {Nolta}, L.~{Page}, D.~N. {Spergel}, M.~{Halpern}, R.~S. {Hill}, A.~{Kogut}, M.~{Limon}, S.~S. {Meyer}, N.~{Odegard}, G.~S. {Tucker}, J.~L. {Weiland}, E.~{Wollack}, and E.~L. {Wright} {\em \apj suppl.} {\bf 192} (2011), no.~2 18, \href{http://arxiv.org/abs/1001.4538}{{\tt arXiv:1001.4538}}.

\bibitem{Planck:2016mks}
{\bf Planck}  {\bf Collaboration} , R.~Adam {\em et~al.} {\em Astron. Astrophys.} {\bf 596} (2016) A108, \href{http://arxiv.org/abs/1605.03507}{{\tt arXiv:1605.03507}}.

\bibitem{Heinrich:2016ojb}
C.~H. Heinrich, V.~Miranda, and W.~Hu {\em \prd} {\bf 95} (2017), no.~2 023513, \href{http://arxiv.org/abs/1609.04788}{{\tt arXiv:1609.04788}}.

\bibitem{Hazra:2018eib}
D.~K. Hazra, D.~Paoletti, F.~Finelli, and G.~F. Smoot {\em \jcap} {\bf 09} (2018) 016, \href{http://arxiv.org/abs/1807.05435}{{\tt arXiv:1807.05435}}.

\bibitem{Millea:2018bko}
M.~Millea and F.~Bouchet {\em \aap} {\bf 617} (2018) A96, \href{http://arxiv.org/abs/1804.08476}{{\tt arXiv:1804.08476}}.

\bibitem{Ahn:2020btj}
K.~Ahn and P.~R. Shapiro {\em \apj} {\bf 914} (2021), no.~1 44, \href{http://arxiv.org/abs/2011.03582}{{\tt arXiv:2011.03582}}.

\bibitem{Qin:2020xrg}
Y.~Qin, V.~Poulin, A.~Mesinger, B.~Greig, S.~Murray, and J.~Park {\em \mnras} {\bf 499} (2020), no.~1 550--558, \href{http://arxiv.org/abs/2006.16828}{{\tt arXiv:2006.16828}}.

\bibitem{aghanim2020planck}
{\bf Planck}  {\bf Collaboration} , N.~Aghanim {\em et~al.} {\em \aap} {\bf 641} (2020) A6, \href{http://arxiv.org/abs/1807.06209}{{\tt arXiv:1807.06209}}. [Erratum: Astron.Astrophys. 652, C4 (2021)].

\bibitem{Tristram:2023:PR4}
M.~Tristram {\em et~al.} {\em \aap} {\bf 682} (2024) A37, \href{http://arxiv.org/abs/2309.10034}{{\tt arXiv:2309.10034}}.

\bibitem{giare2024measuring}
W.~Giar{\`e}, E.~Di~Valentino, and A.~Melchiorri {\em \prd} {\bf 109} (2024), no.~10 103519, \href{http://arxiv.org/abs/2312.06482}{{\tt arXiv:2312.06482}}.

\bibitem{gunn1965density}
J.~E. Gunn and B.~A. Peterson {\em Astrophysical Journal, vol. 142, p. 1633-1636} {\bf 142} (1965) 1633--1636.

\bibitem{fan2006survey}
X.~Fan, M.~A. Strauss, G.~T. Richards, J.~F. Hennawi, R.~H. Becker, R.~L. White, A.~M. Diamond-Stanic, J.~L. Donley, L.~Jiang, J.~S. Kim, {\em et~al.} {\em The Astronomical Journal} {\bf 131} (2006), no.~3 1203.

\bibitem{bosman2022hydrogen}
S.~E. Bosman, F.~B. Davies, G.~D. Becker, L.~C. Keating, R.~L. Davies, Y.~Zhu, A.-C. Eilers, V.~D’Odorico, F.~Bian, M.~Bischetti, {\em et~al.} {\em Monthly Notices of the Royal Astronomical Society} {\bf 514} (2022), no.~1 55--76.

\bibitem{gaikwad2023measuring}
P.~Gaikwad, M.~G. Haehnelt, F.~B. Davies, S.~E. Bosman, M.~Molaro, G.~Kulkarni, V.~D’Odorico, G.~D. Becker, R.~L. Davies, F.~Nasir, {\em et~al.} {\em Monthly Notices of the Royal Astronomical Society} {\bf 525} (2023), no.~3 4093--4120.

\bibitem{Dai:2018nce}
W.-M. Dai, Y.-Z. Ma, Z.-K. Guo, and R.-G. Cai {\em \prd} {\bf 99} (2019), no.~4 043524, \href{http://arxiv.org/abs/1805.02236}{{\tt arXiv:1805.02236}}.

\bibitem{ouchi2020observations}
M.~Ouchi, Y.~Ono, and T.~Shibuya {\em Annual Review of Astronomy and Astrophysics} {\bf 58} (2020), no.~1 617--659.

\bibitem{nakane2024lyalpha}
M.~Nakane, M.~Ouchi, K.~Nakajima, Y.~Harikane, Y.~Ono, H.~Umeda, Y.~Isobe, Y.~Zhang, and Y.~Xu {\em The Astrophysical Journal} {\bf 967} (2024), no.~1 28.

\bibitem{Hu:2003gh}
W.~Hu and G.~P. Holder {\em Phys. Rev. D} {\bf 68} (2003) 023001, \href{http://arxiv.org/abs/astro-ph/0303400}{{\tt astro-ph/0303400}}.

\bibitem{Watts:2020}
D.~J. {Watts}, G.~E. {Addison}, C.~L. {Bennett}, and J.~L. {Weiland} {\em \apj} {\bf 889} (2020), no.~2 130, \href{http://arxiv.org/abs/1910.00590}{{\tt arXiv:1910.00590}}.

\bibitem{Sakamoto:2022nth}
H.~Sakamoto, K.~Ahn, K.~Ichiki, H.~Moon, and K.~Hasegawa {\em \apj} {\bf 930} (2022), no.~2 140, \href{http://arxiv.org/abs/2202.04263}{{\tt arXiv:2202.04263}}.

\bibitem{Mortonson:2007tb}
M.~J. Mortonson and W.~Hu {\em \prd} {\bf 77} (2008) 043506, \href{http://arxiv.org/abs/0710.4162}{{\tt arXiv:0710.4162}}.

\bibitem{Lau:2013zea}
K.~Lau, J.-Y. Tang, and M.-C. Chu {\em Res. Astron. Astrophys.} {\bf 14} (2014) 635--647, \href{http://arxiv.org/abs/1305.3921}{{\tt arXiv:1305.3921}}.

\bibitem{Paoletti:2020ndu}
D.~Paoletti, D.~K. Hazra, F.~Finelli, and G.~F. Smoot {\em \jcap} {\bf 09} (2020) 005, \href{http://arxiv.org/abs/2005.12222}{{\tt arXiv:2005.12222}}.

\bibitem{Gorce:2022cvb}
A.~Gorce, M.~Douspis, and L.~Salvati {\em \aap} {\bf 662} (2022) A122, \href{http://arxiv.org/abs/2202.08698}{{\tt arXiv:2202.08698}}.

\bibitem{jain2023framework}
D.~Jain, T.~R. Choudhury, S.~Mukherjee, and S.~Paul {\em Monthly Notices of the Royal Astronomical Society} {\bf 522} (2023), no.~2 2901--2918.

\bibitem{mukherjee2019patchy}
S.~Mukherjee, S.~Paul, and T.~R. Choudhury {\em Monthly Notices of the Royal Astronomical Society} {\bf 486} (2019), no.~2 2042--2049.

\bibitem{jain2024disentangling}
D.~Jain, S.~Mukherjee, and T.~R. Choudhury {\em Monthly Notices of the Royal Astronomical Society} {\bf 527} (2024), no.~2 2560--2572.

\bibitem{LiteBIRD:2023zmo}
P.~Campeti, E.~Komatsu, {\em et~al.} {\em \jcap} {\bf 06} (2024) 008, \href{http://arxiv.org/abs/2312.00717}{{\tt arXiv:2312.00717}}.

\bibitem{hiramatsu2018reconstruction}
T.~Hiramatsu, E.~Komatsu, M.~Hazumi, and M.~Sasaki {\em Physical Review D} {\bf 97} (2018), no.~12 123511.

\bibitem{blas2011cosmic}
D.~Blas, J.~Lesgourgues, and T.~Tram {\em \jcap} {\bf 2011} (2011), no.~07 034--034, \href{http://arxiv.org/abs/1104.2933}{{\tt arXiv:1104.2933}}.

\bibitem{lidsey1997reconstructing}
J.~E. Lidsey, A.~R. Liddle, E.~W. Kolb, E.~J. Copeland, T.~Barreiro, and M.~Abney {\em Reviews of Modern Physics} {\bf 69} (1997), no.~2 373.

\bibitem{lewis2000efficient}
A.~Lewis, A.~Challinor, and A.~Lasenby {\em \apj} {\bf 538} (2000) 473--476, \href{http://arxiv.org/abs/astro-ph/9911177}{{\tt astro-ph/9911177}}.

\bibitem{tristram2023cosmological}
M.~Tristram, A.~Banday, M.~Douspis, X.~Garrido, K.~G{\'o}rski, S.~Henrot-Versill{\'e}, S.~Ili{\'c}, R.~Keskitalo, G.~Lagache, C.~Lawrence, {\em et~al.} {\em \aap} {\bf 682} (2024) A37, \href{http://arxiv.org/abs/2309.10034}{{\tt arXiv:2309.10034}}.

\bibitem{cen2003universe}
R.~Cen {\em \apj} {\bf 591} (2003) 12--37, \href{http://arxiv.org/abs/astro-ph/0210473}{{\tt astro-ph/0210473}}.

\bibitem{kosowsky2002efficient}
A.~Kosowsky, M.~Milosavljevic, and R.~Jimenez {\em Physical Review D} {\bf 66} (2002), no.~6 063007.

\bibitem{namikawa2010probing}
T.~Namikawa, S.~Saito, and A.~Taruya {\em \jcap} {\bf 2010} (2010), no.~12 027, \href{http://arxiv.org/abs/1009.3204S}{{\tt arXiv:1009.3204S}}.

\bibitem{Hamimeche:2008ai}
S.~Hamimeche and A.~Lewis {\em \prd} {\bf 77} (2008) 103013, \href{http://arxiv.org/abs/0801.0554}{{\tt arXiv:0801.0554}}.

\bibitem{katayama2011simple}
N.~Katayama and E.~Komatsu {\em \apj} {\bf 737} (2011), no.~2 78, \href{http://arxiv.org/abs/1101.5210}{{\tt arXiv:1101.5210}}.

\bibitem{foreman2013emcee}
D.~Foreman-Mackey, D.~W. Hogg, D.~Lang, and J.~Goodman {\em Publications of the Astronomical Society of the Pacific} {\bf 125} (2013), no.~925 306.

\bibitem{Stompor:2008:FGbuster}
R.~Stompor, S.~M. Leach, F.~Stivoli, and C.~Baccigalupi {\em \mnras} {\bf 392} (2009) 216, \href{http://arxiv.org/abs/0804.2645}{{\tt arXiv:0804.2645}}.

\bibitem{Ichiki:2018:delta-map}
K.~Ichiki, H.~Kanai, N.~Katayama, and E.~Komatsu {\em \ptep} {\bf 2019} (2019) 033E01, \href{http://arxiv.org/abs/1811.03886}{{\tt arXiv:1811.03886}}.

\bibitem{Remazeilles:2020:cMILC}
M.~Remazeilles, A.~Rotti, and J.~Chluba {\em \mnras} {\bf 503} (2021) 2478--2498, \href{http://arxiv.org/abs/2006.08628}{{\tt arXiv:2006.08628}}.

\bibitem{Carones:2022:FG}
{\bf LiteBIRD}  {\bf Collaboration} , A.~Carones, M.~Migliaccio, G.~Puglisi, C.~Baccigalupi, D.~Marinucci, N.~Vittorio, and D.~Poletti {\em \mnras} {\bf 525} (2023), no.~2 3117--3135, \href{http://arxiv.org/abs/2212.04456}{{\tt arXiv:2212.04456}}.

\bibitem{Planck:2015}
{\bf Planck}  {\bf Collaboration} , N.~Aghanim {\em et~al.} {\em \aap} {\bf 594} (2016) A11, \href{http://arxiv.org/abs/1507.02704}{{\tt arXiv:1507.02704}}.

\bibitem{Gerbino:2019:likelihood}
M.~Gerbino, M.~Lattanzi, M.~Migliaccio, L.~Pagano, L.~Salvati, L.~Colombo, A.~Gruppuso, P.~Natoli, and G.~Polenta {\em Front. in Phys.} {\bf 8} (2020) 15, \href{http://arxiv.org/abs/1909.09375}{{\tt arXiv:1909.09375}}.

\end{thebibliography}\endgroup

\end{document}